%% file: pipeCDMD.tex
\documentclass{jfm}

\usepackage{graphicx}
\usepackage{newtxtext}
\usepackage{newtxmath}
\usepackage{natbib}
\usepackage{hyperref}
\usepackage{float}
\usepackage{subcaption}
\usepackage{bm}

\newcommand{\rom}[1]
    {\MakeUppercase{\romannumeral #1}}

\hypersetup{
    colorlinks = true,
    urlcolor   = blue,
    citecolor  = black,
}

\newcommand{\RomanNumeralCaps}[1]

\author{Amir Shahirpour\aff{1},
  Christoph Egbers\aff{2}
 \and J\"{o}rn Sesterhenn\aff{1}
 \corresp{\email{joern.sesterhenn@uni-bayreuth.de}}}

\affiliation{\aff{1}Lehrstuhl f\"{u}r Technische Mechanik und Str\"{o}mungsmechanik, Universit\"{a}t Bayreuth, 95440 Bayreuth, Germany
\aff{2}Department of Aerodynamics and Fluid Mechanics, Brandenburg University of Technology Cottbus-Senftenberg}

\begin{document}
\maketitle

\begin{abstract}
   

    Large-scale coherent structures are detected in turbulent pipe flow at $Re_\tau=181$ by having long lifetimes,
    living on large scales and travelling with a certain group velocity. A Characteristic Dynamic Mode Decomposition
    (CDMD) is used to detect events which meet these criteria. To this end, a temporal sequence of state vectors from
    direct numerical simulations are rotated in space-time such that persistent dynamical modes on a hyper-surface are
    found travelling along its normal in space-time, which serves as the new time-like coordinate. Reconstruction of the
    candidate modes in physical space gives the low rank model of the flow. The modes within this subspace are highly
    aligned, but are separated from the remaining modes by larger angles. We are able to capture the essential features
    of the flow like the spectral energy distribution and Reynolds stresses with a subspace consisting of about 10
    modes.  The remaining modes are collected in two further subspaces, which distinguish themselves by their axial
    length scale and degree of isotropy.

\end{abstract}

\begin{keywords}
Authors should not enter keywords on the manuscript, as these must be chosen by the author during the online submission process and will then be added
    during the typesetting process (see \href{https://www.cambridge.org/core/journals/journal-of-fluid-mechanics/information/list-of-keywords}{Keyword
    PDF} for the full list).  Other classifications will be added at the same time.
\end{keywords}


\section{Introduction}
\label{sec:headings}
\input {introduction.tex}

\section{Numerical methods and computational details}
\input {numerical.tex}

\section {Methodology}
\input {methodology.tex}

\section {Results and discussions}
\input {cdmd.tex}

\input {reynolds.tex}
\input {spectra.tex}

\input {isotropy.tex}

\section {Summary and conclusions}
\input {summary.tex}

%
%
%
%

\backsection[Funding]{This joint study was part of the Priority Programme SPP 1881 Turbulent Superstructures of the
Deutsche Forschungsgemeinschaft and funded by grant no. SE 824/33-1 (J.S and A.Sh) and grant no. EG100/24-2 (Ch. E.)}

\backsection[Acknowledgements]{All simulations in this study have been carried out on Norddeutscher Verbund 
f\"{u}r Hoch- und H\"{o}chstleistungsrechnen (HLRN) with project id: bbi00011, using the code of our project partners
within the SPP 1881 (grant no. AV120/3-2).}

\backsection[Declaration of interests]{ The authors report no conflict of interest.}





\bibliographystyle{jfm}
\bibliography{LDMD}

\end{document}

%% file: introduction.tex
Large-scale energetic coherent structures detected in turbulent flows have become an inseparable part of turbulence
studies.  Proof of their existence is promising as it implies that taking advantage of the notion of coherence and
organization, can shed light on the highly dimensional turbulent flows with complex flow patterns.  Coherence in space
and time is commonly known to be caused by flow properties which are maintained by the flow in space and  within a
certain frame of time, so that the maintained property, being for instance a certain type of motion, can be perceived 
as the underlying basis for coherence. 

These structures contribute prominently to the turbulent kinetic energy while diffusing mass and momentum
 and carrying large desirable or undesirable effects such as better mixture or more drag \citep{marusic_2010}. In spite
of large number of studies in the last decade to understand their physical properties and the ease with which they are
spotted by the naked eye, there is still limited consensus in the scientific community on how to define these
structures, what they physically look like, how long they live and how their length scales depend on Reynolds numbers.
It is not fully understood what they feed on, how their regeneration mechanism works and how they interact with each
other or with near wall turbulence.

Three groups of structures are well distinguished in literature by their length scales and wall-normal locations where
they are found. Near wall streaks are known as manifestations of the wall cycle of turbulence and have span-wise spacing
of $\lambda^+ = 100$ \citep{kline_1967}. Their regeneration mechanism has been observed in many studies where their
self-sustainability has been shown. An example would be the study by \citet{jimenez_moin_1991} where a minimal flow unit
is simulated as the smallest channel flow that can maintain turbulence. 

Large scale motions (LSMs) are described as motions whose coherence is maintained as a result of eddies travelling at the
same group velocity \citep{kim_adrian_1999}. Measurements of \citet{bailey_smits_2010} show evidence for existence of
such eddies in the outer layer being detached from the wall with small correlation with the near wall flow, whereas in
the logarithmic region they are more likely to be attached to the wall. This suggests existence of attached LSMs in the
near-wall region and detached ones in the outer layer.  They are known to have stream-wise scale of 2-3 pipe radii and
span-wise length scale of 1-1.5 radii \citep{guala_2006}.  

Very Large-Scale Motions in pipe and channel flow (referred to as VLSMs by Adrian and coworkers) or superstructures in
boundary layer flows (named by Marusic and coworkers), appear to be longer and
have streamwise length scale of 8-20 pipe radii \citep{vallikivi_2015}. While they are mostly seen in the logarithmic
layer in boundary layer flow, they appear in the outer layer of internal flows \citep{monty_etal_2009}.
\citet{kim_adrian_1999} interpret VLSM as a result of stream-wise alignment of LSMs which exist in the outer layer,
whereas \citet{alamo_jimenez_2006} argue that their formation is the result of linear and nonlinear processes.

\citet{toh_itano_2005} consider large-scale structures as part of the turbulence and argue that they feed on their
interactions with the near-wall small-scale structures. \citet{alamo_jimenez_2006b} on the other hand interpret them as
self-sustained structures.  Apart from their regeneration mechanism, many key questions concerning LSM and VLSM are
still unanswered including a uniform scaling law for their identification as well as a clear understanding of their
origin and evolution. Differing views on the origin and nature of low wave number VLSM question their dependence on
geometry and outer layer variables. 

Spectral analysis has been one of the key approaches commonly used to learn about the properties of such structures.
Their foot prints can be followed by observing the premultiplied velocity spectra which represent the energy
distribution in the wave number space. At sufficiently large Reynolds numbers two peaks appear in contour plots of
spectra which are associated with VLSM and LSM \citep{rosenberg_2013}. The signature of large-scale energetic structures
are hereby followed and  their length scales and energy content at different wall normal positions are determined.
Taking advantage of this signature, \citet{bauer_2019} apply a two dimensional Fourier cut-off filter to separate the
structures based on the their known length-scales to investigate which length scales are responsible for feeding the
largest scales and which ones feed from them. Besides differing views on the nature and origin of turbulent structures,
the suitable approach for their analysis is also still under debate. Following the spectral peaks helps to follow foot
prints of structures, but cannot provide insight to their evolution and interactions. 

One of the major difficulties arising while studying the physical properties of large-scale coherent structures, is that
many of the findings can be biased by influences of smaller-scale structures and instabilities. This has led to an
increasing interest in extracting the structures from the turbulent flows and to study their properties in absence of
small-scale structures. The latter, together with recent availability of large numerical and experimental datasets has
led to increasing popularity of data driven methods. 

After introduction of Proper Orthogonal Decomposition (POD) to fluid dynamics by \citet{lumley_1967}, numerous
variations of this method were proposed building on the main idea which was to extract spatial and temporal flow
structures from numerical and experimental data by decomposing the flow to spatially uncorrelated modes.  This was
particularly desirable as the largest amount of energy could be captured with the fewest number of modes, but also
required  the flow to be projected to orthogonal basis, hence removing the possibility for the modes to linearly
interact. An example would be the study by \citet{hellstrom_smits_2014} who applied snapshots POD \citep{sirovich_1987}
to cross-sectional PIV measurements, and found that the first 10 snapshots POD modes contribute 43\% to average Reynolds
shear stress and 15\% to the kinetic energy. In a different approach, Dynamic Mode Decomposition (DMD) was introduced by
\citet{schmid_sesterhenn_2008} decomposing the flow to correlated spatial modes possessing certain temporal frequencies
and decay rates. 

Majority of these methods decompose the flow on a stationary frame of reference  leading to the need for large number of
modes to describe the convecting features in the transport-dominated flows. This issue is addressed by several studies
(\citet{rowley_marsden_2000} and \citet{reiss_2018}) introducing a spatial transformation in form of a shift.
\citet{sesterhenn_shahirpour_2019} proposed a different approach by applying a spatio-temporal transformation in form of
a rotation in space and time on a moving frame of reference along the characteristics of the flow. Hereby, they observed
a faster drop of singular values compared to the shifted reference frame. In what follows we apply a CDMD to DNS data of
turbulent pipe flow.


%% file: numerical.tex
The data used for the study is generated using an open-source, hybrid parallel DNS code \citep{lopez_2020}. Hereby,
Navier-Stokes equations are solved in cylindrical coordinates for an incompressible pipe flow fulfilling mass and
momentum conservations given by 

\begin{equation}
  \nabla \cdot \mathbf{u} =0 \, ,  
    \quad \partial_t \mathbf{u}+\mathbf{u} \cdot \nabla \mathbf{u}=- \nabla p + \frac{1}{Re_b}\nabla^2 \mathbf{u},
\label{NSeq}
\end{equation}

\noindent where $\mathbf{u}(\mathbf{x},t)$ and $p(\mathbf{x},t)$ represent velocity field ($u, v, w$)  in cylindrical
coordinates $\mathbf{x} = (x,r,\theta)$ and dynamic pressure respectively. The governing equations are solved for
velocity and pressure, being discretised with a combined Fourier-Galerkin / Finite Difference method in space and using
a semi-implicit fractional-step of \citet{hugues_1998}, using second-order-accurate backwards differences and second
order linear extrapolation for nonlinear term. More details on the numerical scheme can be found in the study by
\citet{shi_2015}. Simulations are carried out at bulk Reynolds number of $Re_b = 2R U_b/\nu =5300$ for pipe length of
$L=50R$ with $R$, $U_b$ and $\nu$ being respectively the pipe radius, bulk velocity and kinematic viscosity.  After the
final grid refinement, calculations have been advanced for 400 convetive time steps $t_c=R/U_b$, during which $
\textup{CFL}_{max} = 0.5$ was maintained, leading to simulation time step of $dt = 4.93\times 10^{-4} \, t_c $. The grid
spacing measured in wall units is chosen so that there are 5 and 20 points bellow $y^+=1$ and $y^+=10$ respectively,
with the first point in the vicinity of the wall at $y^+=0.026$. The $+$ superscript denotes normalisation by inner
scaling using viscous length-scale $\nu/u_{\tau}$ and friction velocity $u_{\tau} = \sqrt{\tau_w/\rho}$ where $\tau_w$
and $\rho$ are the wall shear stress and density respectively.  Further details on the simulation and grid spacing are
mentioned in table \ref{tab:DNS}. Results are validated by comparing the statistical flow properties with benchmakr DNS
data in the next chapters.

\begin{table}
  \begin{center}
  \def~{\hphantom{0}}
	\begin{tabular}{@{}cccccccc@{}}
            $Re_{\tau}$ & $u_{\tau}/U_b$ & $N_x \times N_r \times N_{\theta} $  & $(\Delta x^+)$ & 
            $(\Delta r^+)_{min}$ & $(\Delta r^+)_{max}$ & $(\theta \Delta r^+)_{min}$ & $(\theta \Delta r^+)_{max}$\\[5pt]
            181   &	0.068       & 1800 $\times$ 120 $\times$ 286   & 5&0.08 &2.2 &0.002 &0.04 \\

	\end{tabular}
  \caption{Details of the simulation and grid spacing. $N_x$, $N_r$, and $N_\theta$ are the number of grid points in
  $(x, r, \theta)$ directions and $Re_{\tau} = Ru_{\tau}/\nu$ is defined as shear Reynolds number.}
  \label{tab:DNS}
  \end{center}
\end{table}


%% file: methodology.tex
\subsection{Characteristic DMD}

Investigating transport-dominated phenomena on a stationary frame of reference adversely influences the
observations. To remedy this issue, \citet{sesterhenn_shahirpour_2019} proposed a Characteristic DMD. The essence of a
characteristic decomposition of the flow is to seek coherence as a persistent behaviour observed in space and
time coupled together on a moving frame of reference, as opposed to spatial or temporal coherence individually. 
They introduced a transformation $\mathcal{T}$ in form of a rotation in space and time 
$\mathcal{T} (\mathbf{u} (x,r,\theta,t)) = \mathbf{u}(\xi,r,\theta,\tau)$ and used the drop of singular values as a
measure of how well the convected phenomena can be described on each frame of reference. 

Two major advantages were
presented for the spatio-temporal transformation. The first one is that convected phenomenon could be described on the
rotated frame with far fewer modes compared to a stationary frame. In addition, it was shown that singular values drop
faster along the characteristics compared to those taken on a shifted moving frame which is obtained by a purely spatial
transformation. The second advantage is that as expected, dynamics of the detected structures are captured more
accurately.

Having chosen the frame of reference, the decomposition method is selected based on the fact that the goal of this study
is to analyse the interactions between the modes. We intend to present a framework in which the origins of structures,
their regeneration mechanism, their sustainability and finally their decay process can be investigated. Therefore, the
obtained eigen modes should be found such that they can give energy to other modes or to feed from them, and as a
result, should not be forced to be normal to each other. To this end, the standard dynamic mode decomposition
\citep{schmid_2010} has been taken as the main basis for decomposing the flow field. Three subsets of the modes are
detected, reconstructed in spatio-temporal space and transformed back to physical space, where their contributions to
Reynolds stress tensor and their anisotropy invariant maps are studied.  Further details of the method can be found
in the relevant manuscript.

A reference is needed to validate the identity of captured structures. What many studies have in common in their
definition of coherent structures, is the foot print they leave behind in the Fourier space, in premultiplied energy
spectra. Therefore, we verify our detected structures, by how well they represent the spectral peak and therefore, we
use velocity field as the state vector in our analysis.

%% file: cdmd.tex
\subsection{Direction search}

The main goal in the first step is to find the direction of characteristics along which the large-scale features of the
flow can be described with fewest modes possible.  The slope of the characteristics represents the group velocity $u_g$
at which the large-scale features are being convected and is defined as the axial length-scale travelled per unit
convective time  defined as $t_c = R/U_b$.  

\begin{figure}
\centering

    {{\includegraphics [scale=0.95] {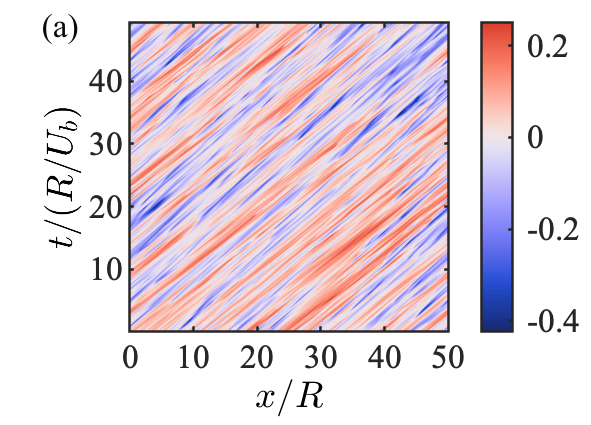} }}
    {{\includegraphics [scale=0.95] {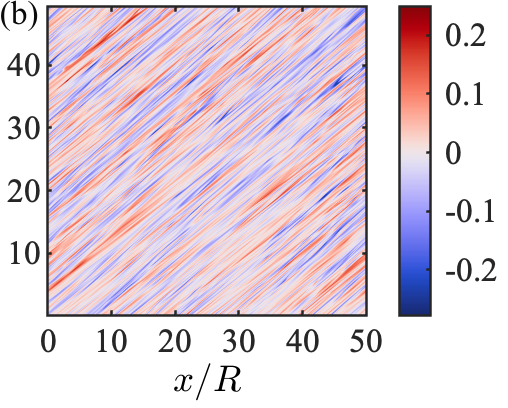} }}
    {{\includegraphics [scale=0.95] {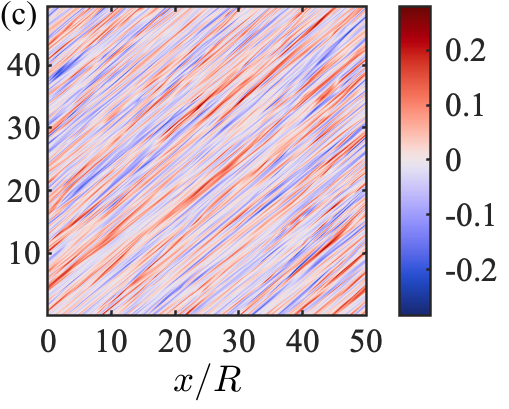} }}
\caption{Space-time diagrams for $u$ (a), $v$ (b) and $w$ (c) normalised by bulk velocity $U_b$ at
    wall-normal location $y/R= 0.5$.}
\label{spaceTime}
\end{figure}

In figure \ref{spaceTime}, space-time diagram is shown for three velocity components at wall-normal location $y/R=0.5$
for one azimuthal location.  The colourmap represents the corresponding velocity component normalised by bulk velocity.
Although several group velocities can be observed for each component, one dominant group velocity can be perceived which
corresponds to the energetic large-scale events.  The dominant group velocity will get essentially smaller by moving
closer to the wall and into the wall layer, and it will be larger in the outer layer and close to the pipe axis.  A
second observation is that the main group velocity appears to stay relatively constant for 50 $t_c$ which is the time
required to travel through the pipe once.

\begin{figure}
\centering
   {{\includegraphics{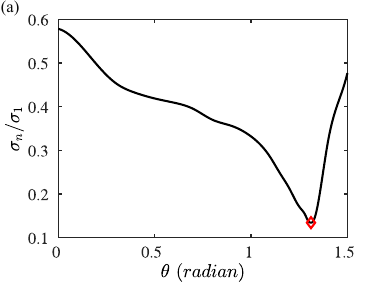} }}
   {{\includegraphics{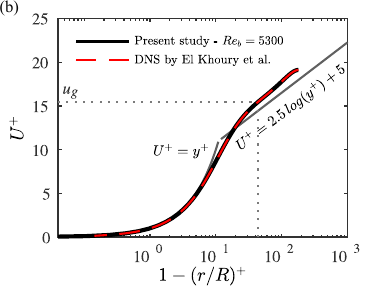} }}
    \caption{Drop of singular values for a range of rotation angles (a) and mean radial velocity profile compared
    against
    the benchmark data (b).}
\label{svDrop}
\end{figure}

The objective is to decompose the flow into modes which describe the complete velocity field. Therefore, the direction
along which the decomposition is applied, should be chosen optimally for all velocity components. Optimality here is
defined by detection of large-scale features using a minimal number of modes and is quantified by the drop of singular
values along the characteristics. For each time-step the entire velocity field is stacked in one column vector which
forms one of the columns of matrix $M_{(N_{ph} \times N_t)}$ with $N_{ph}$ and $N_t$ corresponding to the number of
spatial points in physical space and time-steps respectively.  The spatio-temporal rotation is then applied to $M$ for a
range of angles spaced $0.1$ radian from each other.  After each rotation, a singular value decomposition is carried out
and the drop of singular values are recorded as shown in figure \ref{svDrop}a. A piecewise cubic interpolation is then
used to fit a curve to all the points and to find the maximum drop which is shown with a red marker for rotation angle of
$\theta_g = 1.311$ corresponding to group velocity of $u_g = 1.06 \, U_b = 15.5 \, u_{\tau}$.  This group velocity
is equal to the mean radial velocity found at wall-normal location $y^+ = 1-(r/R)^+ = 44$.  In figure
\ref{svDrop}b,  $u_g$ is annotated along with the mean radial velocity profile compared with the benchmark data by
\citet{el-khoury_schlatter_2013}. By rotating the matrix $M$ by $\theta_g$, the data will be transformed to a moving frame of
reference with the direction of characteristics serving as the new time coordinate.  We search for coherent structures
in planes normal to the characteristics as they travel in space and time and undergo minimal changes while maintaining
their coherence. 

\subsection{Decomposition and subspace detection}

\begin{figure}
\centering
\centering \includegraphics[width=0.9\linewidth]{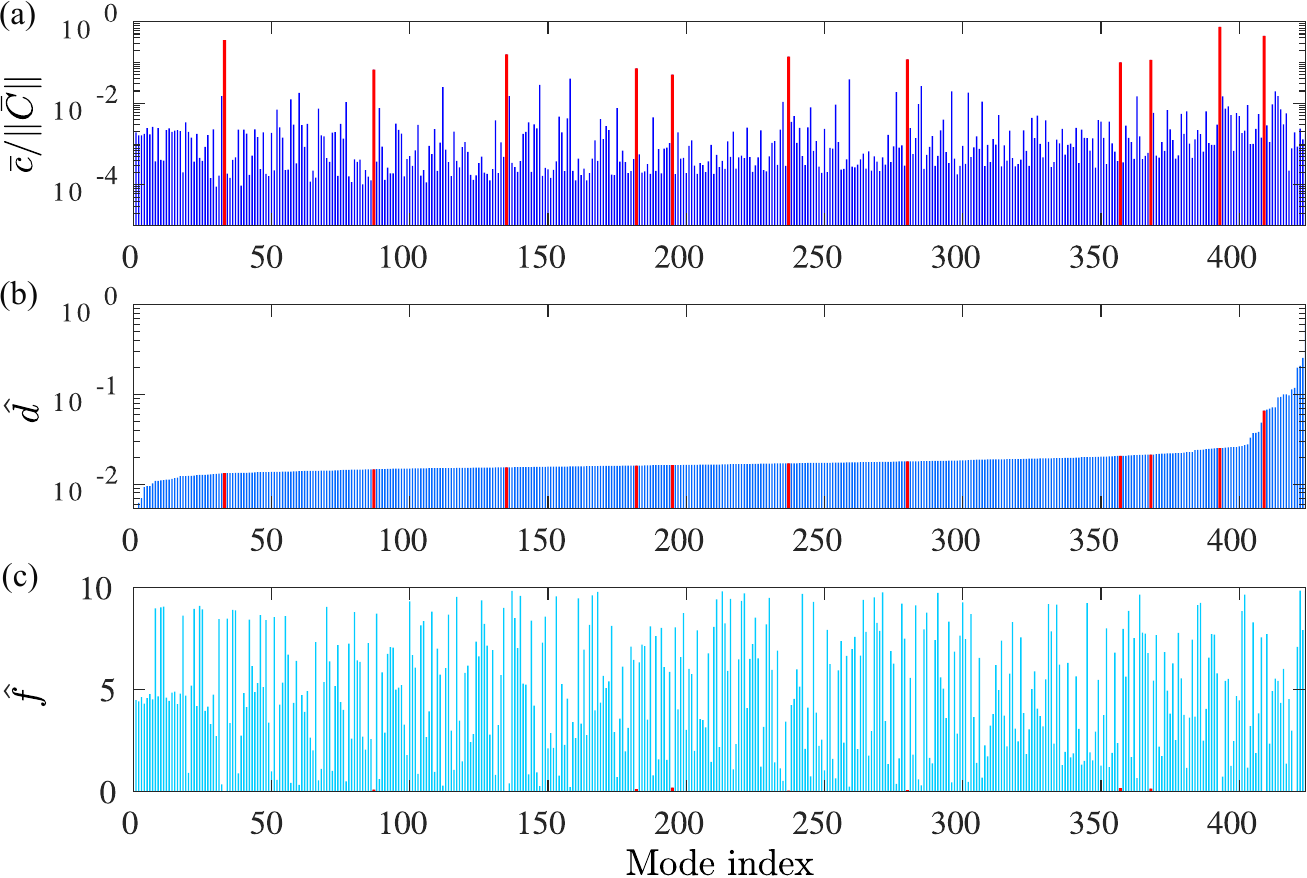} 
\caption{Dynamic mode amplitudes, decay rates and frequencies.}
\label{spctrm}
\end{figure}

Having detected the optimal group velocity, matrix $M$ is formed using 500 timesteps with spatial resolution of
$(900\times 60 \times 143)$ in $(x, r, \theta)$ directions. To ensure that the dynamics of the modes are captured
correctly, timestep of $dt_{\cal{C} \text{DMD}} = 0.2 \, t_c$ is chosen between the columns of $M$. Therefore, each
event moving at $U_b$ propagates two times through the entire pipe. Transforming the data to spatio-temporal space and
choosing the largest $\xi - \tau$ window in the rotated frame of reference results in the snapshots matrix in
spatio-temporal space $X_{st} = \mathcal{T}(M)$, with $N_{\xi} = 290$ and $N_{\tau} = 843$ points along $\xi$ and $\tau$
respectively. A standard DMD is carried out to decompose $X_{st}$ into the dynamic modes $\boldsymbol{\phi}_i$ and their
corresponding coefficients $c_i(\tau)$ such that $X_{st} = \varPhi C$ where $\varPhi$ and $C$ are matrices of dynamic modes
and their coefficients for all timesteps. Continuous-time eigenvalues are transformed back to physical space with their
real and imaginary parts representing decay rates and frequencies of the modes respectively in physical time. In figure
\ref{spctrm}, time averaged mode coefficients normalised by their $\mathcal{L}_2$ norm, dimensionless decay rates $\hat{d}=
d/(U_b/R)$ and frequencies $\hat{f}=f/(U_b/R)$ are plotted with the modes being sorted by their decay rates. All the
frequencies in spatio-temporal space are within the range of $0 \leq \hat{f}_{st} < 2.5 $ which corresponds to 
$0 \leq \hat{f} <10 $ after transformation to physical space.

Next, a subset of modes is to be selected constituting a subspace (subspace \rom{1}) to fulfil certain criteria which
are chosen with the knowledge, that for turbulent pipe flow at this Reynolds number, there exists only one peak in the
premultiplied energy spectra. The first criteria is that subspace \rom{1} should accommodate energetic structures with
large spatio-temporal length-scales.  Therefore, it is expected to have a large contribution to the spectral peak which
is known as the footprint of large-scale structures in premultiplied energy spectra. Given the nature of coherent
structures, the second criteria dictates that the modes in this subspace should not possess large decay rates.  This is
to ensure that energetic modes with short lifetimes will not be a member of this subspace. Similarly, the candidate
modes are expected to have small frequencies and not undergo strong oscillations. 

We hypothesize that the modes fulfilling the mentioned criteria are expected to possess another significant property.
Due to the spatio-temporal coherence of the flow captured by these modes, they are expected to have major interactions with
each other, but have smaller interactions with the rest of the modes. We define this interaction in terms of the angle
between the modes as well as the energy which is gained or lost by the flow as a result of presence of each two modes in a
subspace. This implies that the modes in subspace \rom{1}, besides being energetic, should have small angles between
each other and large ones with the remaining modes.

To calculate the energy of a subspace, we first consider subspace $S$ comprised of two modes $S = \{\bm{\phi}_1,\bm{\phi}_2\}$ and
coefficients matrix $C_S$ with rows defined as $c_1(\tau)$ and $c_2(\tau)$ that can be used to obtain $X_S = S C_S$.
Columns of $X_S$ and $C_S$ can be used to write for each timestep $\boldsymbol{\chi}_S(\tau) = S \, \bm{c}_S(\tau)=
\bm{\phi}_1 c_1(\tau) + \bm{\phi}_2 c_2(\tau)$. The total energy of $S$ integrated along $\tau$ is then defined by

\begin{equation}
    E_S = \sum_{\tau = 1}^{N_{\tau}} \bm{c}_S^*(\tau) \, S^* S \, \bm{c}_S (\tau),
\label{eqE_Q}
\end{equation}

\noindent and energy of $S$ can be written for each time step $\tau$ as

\begin{equation}
\begin{split} 
    E_{S}(\tau) & =  \chi_S^*(\tau) \chi_S (\tau)  = \bm{c}_S^*(\tau) \, S^* S \, \bm{c}_S(\tau) =
    \Bigl( c_1^*(\tau) {\bm{\phi}^*_1} + c_2^*(\tau) {\bm{\phi}^*_2} \Bigr) \Bigl( {\bm{\phi}_1}c_1(\tau) 
    + {\bm{\phi}_2}c_2(\tau) \Bigr) \\
      & = \underbrace{ c_1^*(\tau) {\bm{\phi}^*_1}{\bm{\phi}_1}c_1(\tau) }_{E_1(\tau)}  + 
          \underbrace{ c_1^{*}(\tau) \bm{\phi}_1^* \bm{\phi}_2 c_2(\tau) 
          + c_2^{*}(\tau) \bm{\phi}_2^* \bm{\phi}_1 c_1(\tau)}_{E_{1|2}(\tau)} +
          \underbrace{ c_2^{*}(\tau) \bm{\phi}_2^*{\bm{\phi}_2}c_2 (\tau) }_{E_2(\tau)}.
\end{split}
    \label{subEn}
\end{equation}


The terms $E_1(\tau)$ and $E_2(\tau)$ in equation \ref{subEn} correspond to the energy of modes $\bm{\phi}_1$ and
$\bm{\phi}_2$ respectively at one timestep, and the term $E_{1|2}(\tau)$ represents the energy added to or taken from
$X_s$ as a result of interaction between $\bm{\phi}_1$ and $\bm{\phi}_2$.  For modes that are orthogonal to each other,
the term $E_{1|2}$ vanishes and for mode pairs with small angles, $E_{1|2}$ can have large positive or negative values.
Equation \ref{subEn} can be generalised to the case where $\bm{\phi}_1$ and $\bm{\phi}_2$ are each separate subspaces. 

\begin{figure}
\centering
     { \includegraphics{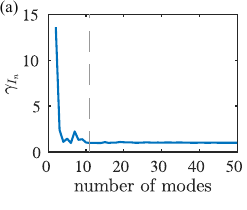} }
     { \includegraphics{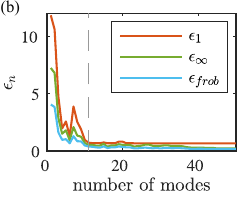} }
     { \includegraphics{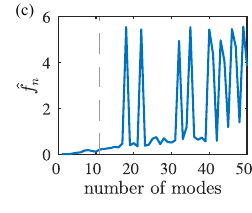} }
    \caption{Cumulative energy (a), relative error (b) and frequencies (c) of the first n dominant modes.}
\label{enI-Err-fr}
\end{figure}

To detect a subspace $\text{I}_n$ with $n$ most energetic modes that represents the full-field energy with fewest number
of modes, and to observe how the subspace energy changes as the next energetic mode is added to it, cumulative energy is
calculated for the first $n$ dominant modes, integrated along $\tau$ and normalised by the total energy as

\begin{equation}
    \gamma_{I_n} = E_{I_n}/E_{\varPhi}.
\label{cumEn}
\end{equation}

\noindent and plotted in figure \ref{enI-Err-fr}a for the first 50 modes. $E_{I_n}$ represents energy of subspace \rom{1}
possessing $n$ modes integrated over time. A fast drop is observed for the first few
modes added, where two minima are observed for 4 and 6 modes resulting in subspace energy close to 1. 
($\gamma_{I_4} = 1.1$ and $\gamma_{I_6} = 0.9 $). Adding more modes increases the energy, but finally by having 11
modes, subspace energy will drop again to $\gamma_{I_{11}} = 0.9$. It is clear that adding the next modes makes only minimal
changes in the subspace energy. 

Next, relative error is calculated for reconstruction of modes in subspace $\text{\rom{1}}_n$ having $n$
modes with the corresponding coefficients matrix $ C_{I_n}$ (equation \ref{relErr}). Three matrix norms have been used 
with $p = \{1, \infty,F\}$ for one-norm, infinity norm
and frobenius norm respectively. 

\begin{equation}
    \epsilon_{n} = \| X - I_n C_{I_n} \|_p  / \| X \|_p .
\label{relErr}
\end{equation}

As shown in figure \ref{enI-Err-fr}b, all relative errors reach two minima for 4 and 6 modes, increase for 7 modes, 
and then drop strongly for 11 modes while minimally changing beyond that point. As depicted in figure \ref{enI-Err-fr}c, 
these 11 modes have very small frequencies compared to the rest of the modes.

The candidate 11 modes are highlighted in red in figure \ref{spctrm} with red bars. They have a small mean decay rate of
$\hat{d_I} = 0.022$ and undergo minimal oscillations with average frequency of $\hat{f_I} = 0.086$ in the range of $0
\leq \hat{f} \leq 0.2$. All the remaining modes oscillate with larger frequencies $0.22 \leq \hat{f} \leq 9.86$ with
the exception of mode 420 which has a large decay rate and small amplitude and therefore does not meet the criteria to
be part of this subspace. 8 of the candidate modes have large amplitudes $ \bar{c}/\| \bar{C} \| \, \geq 0.1$  and
the rest, in spite of having smaller amplitudes $0.05 \leq \bar{c}/\| \bar{C} \| \, \leq 0.07$, still possess much
smaller frequencies compared to the rest of the modes.  Therefore, based on the cumulative energy of subspace \rom{1},
its relative error, mode amplitudes, their decay rates and frequencies, the first 11 dominant modes are chosen as
members of subspace \rom{1}. 

\begin{figure}
\centering
     { \includegraphics{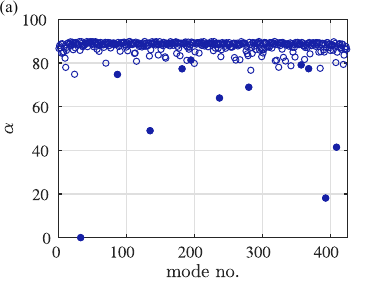} }
     { \includegraphics{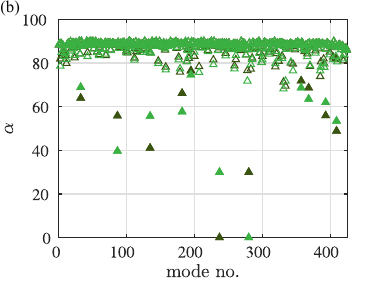} }
\caption{Angles between mode-pairs $m_{33} \angle m_i$ (a), $ m_{237} \angle m_i$ and $m_{280} \angle m_i$ (b).}
\label{modesAngle}
\end{figure}

Having detected a subset of energetic modes matching the mentioned criteria, we verify orthogonality of each member of
this subset to modes residing inside and outside the subset. Mode-pair angles $m_{33} \angle m_i$ are plotted as an
example in figure \ref{modesAngle}a with blue markers showing the angles that mode 33 (one of the members of subspace
\rom{1}) makes with all the other modes. Filled blue markers, correspond to subspace \rom{1} members. It is readily seen
that majority of the modes outside subspace \rom{1}, are almost orthogonal to mode 33 as they accumulate close to
$\alpha = 90$. On the other hand, the smallest angles are made with members of subspace \rom{1} ($m_{393}$ and
$m_{409}$), indicating interaction between $m_{33}$ with a small decay rate, with two modes with rather larger decay
rates. In figure \ref{modesAngle}b, mode pair angles $m_{237} \angle m_i$ and $m_{280} \angle m_i$ are plotted. Here it
is also observed that modes outside subspace \rom{1} are mostly orthogonal to $m_{237}$ and $m_{280}$. These two modes
appear to make small angles with one another and rather larger angles with the rest of the modes in subspace \rom{1}. A
similar behaviour exists for all the modes in subspace \rom{1}.

\begin{figure}
\centering
     {{ \includegraphics{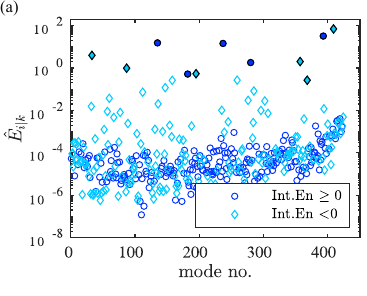} }}
     \hspace{4mm}
     {{ \includegraphics{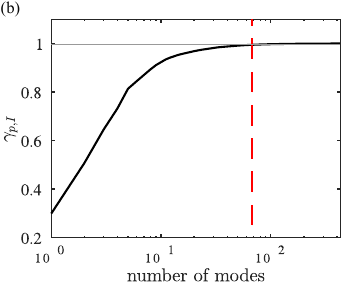} }} 
    \caption {Normalised energy interactions between mode $m_{237}$ and all the other modes (a), and cumulative energy
    interaction between subspace \rom{1} and the rest of the modes (b).}
\label{cumEnInt}
\end{figure}

A small angle between two modes, provides the potential for a large energy interaction. But as inferred from equation
\ref{subEn}, the term $E_{1|2}$ is dependant also on the mode coefficients besides the inner product of the two modes.
Therefore in the next step, integrated energy interactions $\hat{E}_{i|k}$ are calculated between each mode ($k$) in
subspace \rom{1} and all the other modes ($i$) normalised by the total energy of the flow (with $\hat{.}$ denoting the
normalisation). The results are plotted for $\hat{E}_{i|237}$ in figure \ref{cumEnInt}a with circles and diamond markers
corresponding to positive and negative values respectively. Filled markers represent modes in subspace \rom{1}, which
show clearly the largest interactions with $m_{237}$, some with positive and some with negative values. Apart from the
contributions of the modes in subspace \rom{1} (filled markers), two distinct regions also appear in this plot. A
smaller number of modes can be seen at $\hat{E}_{i|237} \geq 10^{-3}$ and majority of them seem to be accumulated below
this limit. This implies that there are certain modes outside subspace \rom{1}, which are interacting more than the rest
with $m_{237}$. These two regions appear for all the modes in subspace \rom{1} indicating emergence of a second
subspace, whose members are chosen based on how much energy they bring or take from the flow while interacting with
subspace \rom{1}. To detect the modes fitting in the new subspace, the term $\hat{E}_{i|I}$ should be calculated for all
the members of subspace \rom{1} as

\begin{equation}
    \hat{E}_{i|I} = \frac{\sum\limits_{\tau=1}^{N_{\tau} } \, 
                          \sum\limits_{k=1}^{N_I} 
                          \Bigl( c^*_i(\tau) \, \bm{\phi}^*_i \bm{\phi}_k \, c_k(\tau) + 
                          c^*_k(\tau) \, \bm{\phi}^*_k \bm{\phi}_i \, c_i(\tau) \Bigr) }
    {\sum\limits_{\tau=1}^{N_{\tau}} \bm{c}^*(\tau) \, \varPhi^* \varPhi \, \bm{c}(\tau) },
\label{eqEnInt}
\end{equation}


\noindent with $N_I$ being the number of modes in subspace $I$, and $\tau_i$ and $\tau_e$ being the initial and last time
step along $\tau$ respectively. The vector calculated using equation \ref{eqEnInt} is sorted in a descending order and 
modes in subspace \rom{1} are excluded from the set in order to detect the largest contributions to subspace \rom{1}.
Cumulative energy interaction is then given for the first $p$ dominant contributions by

\begin{equation}
    \gamma_{p,I} = \frac{ \sum\limits_{j=1}^{p} \hat{E}_{j|I} }{  \sum\limits_{j=1}^{N-N_I} \hat{E}_{j|I} }, 
\label{eqCumEnInt}
\end{equation}

\noindent with $N$ being the total number of modes, and is plotted in figure \ref{cumEnInt}b. Cumulative energy
contribution rises rapidly with the first 10 modes and reaches a saturation point after 60 modes, beyond which the
energy does not change much by adding the remaining modes. Taking 67 modes, where a red dashed line is plotted,
captures 98\% of the total contribution (grey solid line). 

Having detected a second subspace, the remaining modes are grouped together to form the third subspace.  Subspaces
\rom{1}, \rom{2} and \rom{3} each with 11, 67 and 346 modes amount to 3\%, 15\% and 82\% of the total number of modes
respectively. Their total kinetic energy is calculated using equation \ref{eqE_Q} being equal to $97\%$, $15\%$ and
$2\%$ of the snapshots energy for the first, second and third subspaces respectively.  Subspace interactions \rom{1} $|$
\rom{2} and \rom{2} $|$ \rom{3} lead to $12\%$ and $2\%$ energy loss, whereas interactions between the first and third
subspaces does not cause any overall energy gain or loss.

\begin{figure}

\centering
     {{ \includegraphics{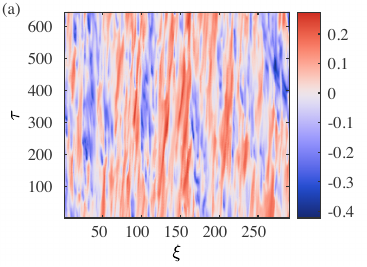} }}
     {{ \includegraphics{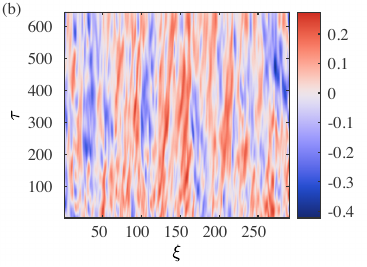} }}\\
     {{ \includegraphics{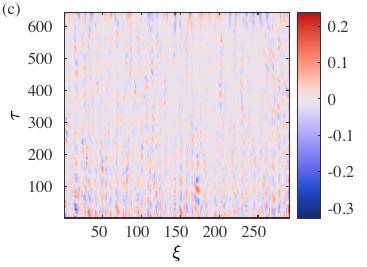} }}
     {{ \includegraphics{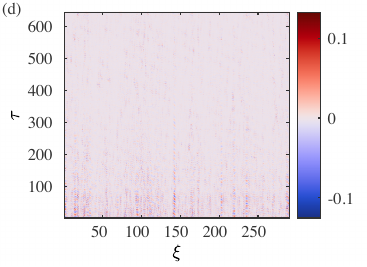} }}
    \caption{Space-time diagram in spatio-temporal space for the full-field (a) and subspace \rom{1} (b), \rom{2} (c)
    and \rom{3} (d), at wall-normal location $y=0.34R$ and one azimuthal point}
\label{subST}
\end{figure}

Each subspace is then reconstructed along $\tau$. To have a visual comparison between the full-field and the subspaces,
space-time diagrams are plotted for axial velocity components in figure \ref{subST} at radial location $y=0.34 R$ ($y^+=
61.5$) for one azimuthal location. Comparing the full-field with Subspace \rom{1} in figures \ref{subST}a and \ref{subST}b,
it can be seen that the large-scale flow patterns are present very well in spite of the fact that only 3\% of the modes
exist in this subspace. Magnitudes of negative and positive perturbations agree well with those of the full-field.
Small-scale patterns are clearly missing from the reconstruction as expected. Dominant structures in this subspace 
appear to remain stationary along the direction of $\tau$. 

Subspace \rom{2} in figure \ref{subST}c accommodates small-scale patterns with perturbations which are considerably less
energetic than those captured in subspace \rom{1}. Oblique patterns emerging show that structures here have different group
velocities compared to the dominant one. Some appear to move backwards relative to the moving frame of reference
indicating a slower convection velocity, whereas others move forward at higher velocity.  Absence of strong vertical
patterns in this figure shows that no energetic structure moving with the dominant group velocity is present in subspace
\rom{2}.

Subspace \rom{3} with $82\%$ of the modes bears traces of some small-scale patterns similar to those in subspace \rom{2}, but is
mainly populated with very small-scale structures. No dominant group velocity is observable in this subspace.


\subsection{Subspaces in physical space}

\begin{figure}
\centering
     \centering \includegraphics{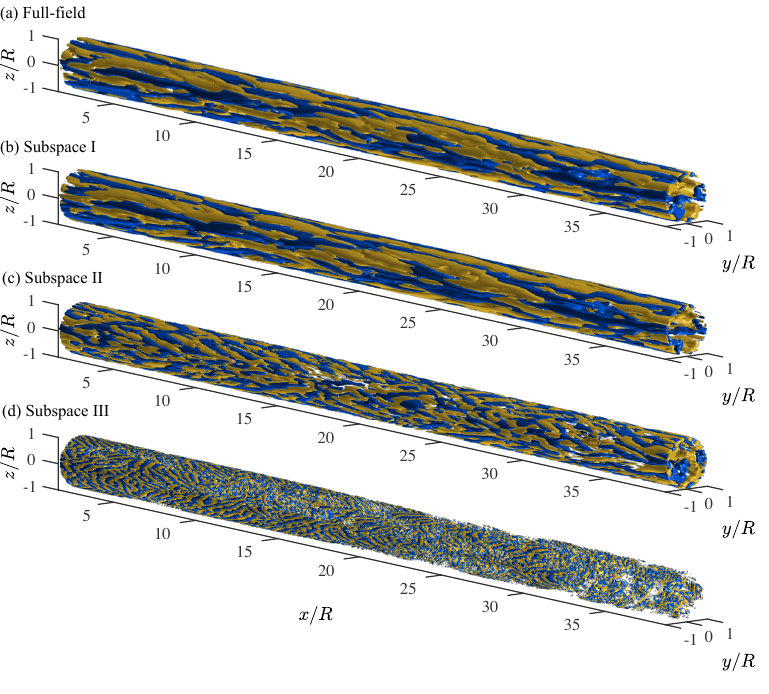}
\caption{Iso-surfaces of axial velocity component of the full-field compared against each subspace. Iso-levels for the
    full-field (a) and subspace \rom{1} (b) are identical with yellow and blue corresponding to $u = \pm 0.1 \, U_b$. In
    subfigures (c) and (d) iso-levels of $u= \pm 0.04 \, U_b$ and $\pm 0.005 \, U_b$ are plotted respectively for subspaces
    \rom{2} and \rom{3}.}
\label{ux3D}
\end{figure}

Each reconstructed subspace is transformed back to physical space. In figure \ref{ux3D}, iso-surfaces of streamwise
velocity component is shown for the full-field (figure \ref{ux3D}a) and for each subspace. Similar to what was seen in
the spacetime diagram, the structures in subspace \rom{1} (figure \ref{ux3D}b) appear to be similar to those in
full-filed.  This resemblance is observed in terms of where high and low momentum regions are located and also in terms
of amplitudes of perturbations (In both subfigures iso-levels $u = \pm 0.1 \, U_b$ are plotted).  Axial
length-scales in both figures perceived from the large-scale structures agree well and they will be examined in the next
chapters in premultiplied spectra. 

Subspace \rom{2} in figure \ref{ux3D}c on the other hand, accommodates only smaller scale structures with lower
perturbation magnitudes (with iso-levels $u = \pm 0.04 \, U_b$). The modes in this subspace were chosen based on the
level of their interactions with subspace \rom{1} causing large energy gains or losses.  On the other hand it was shown
in figure \ref{enI-Err-fr} that the total energy of the flow will not change drastically beyond 11 modes. This implies
that although the two subspaces have large energy interactions, the overall energy of subspace \rom{1} remains relatively
constant. Subspace \rom{3} is plotted with iso-levels $u = \pm 0.005 \, U_b$ with two major length-scales being
present in the flow, both of which are smaller than those present in the other subspaces. 

\subsection{limitations and constraints}
\label{sec:limitations}

Due to two reasons the statistical turbulence properties of the full-field diverge from those of the snapshots matrix in
physical space $X_{ph} = \mathcal{T}^{-1}(X_{st})$. The first reason is that in order for the second order statistics to
converge, 4000 data realisations recorded for 400 convective timesteps have been used, whereas taking the same number of
timesteps for DMD was not possible due to memory limitations. The second reason is the linear interpolation used for the
spatio-temporal transformations. 

The first limitation could be only partially removed using a streaming DMD \citep{hemati_rowley_2014} and at the expense
of truncating the singular values. As in this study it was intended to keep all non-zero singular values, a streaming
DMD was not used. Employing higher order interpolation schemes and using larger number of timesteps, substantially
increase the computation time specially at higher Reynolds numbers. It was also observed that the present setup does not
bias the conclusions. Therefore, turbulence properties of the subspaces in subchapters \ref{sec:reynolds} and
\ref{sec:isotropy} are presented using three references. The first two references are the full-field and DNS data by
\citet{el-khoury_schlatter_2013} which are compared against each other to validate the simulation results.  $X_{ph}$ serves as the third reference against
which the subspaces are compared. In subchapter \ref{sec:spectra}, the difference between the length scales in snapshots
and the full-field is compensated by applying the same correction to the snapshots and all subspaces.

%% file: reynolds.tex
\subsection {Contribution to Reynolds stress tensor}
\label{sec:reynolds}

\begin{figure}
\centering
    {{ \includegraphics{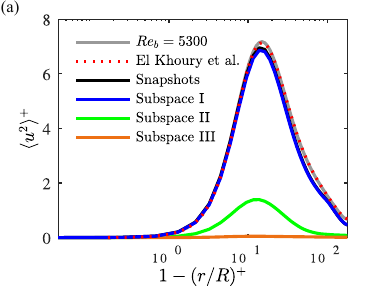} }}
    {{ \includegraphics{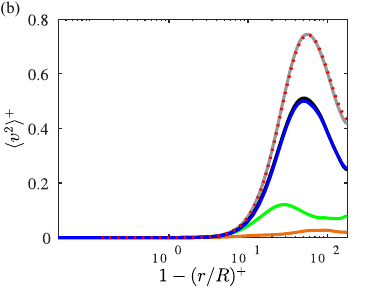} }}\\
    {{ \includegraphics{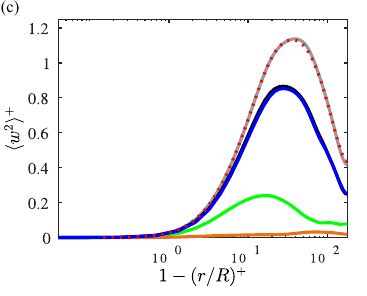} }}
    {{ \includegraphics{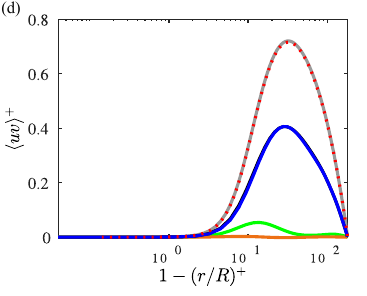} }}
\caption{Reynolds stress components of the full-field compared against those of the benchmark data, each subspace and
    the snapshots matrix}
\label{fig:reyStress}
\end{figure}

Contributions of each subspace to components of Reynolds stress tensor are calculated in physical space and are compared
against the snapshots $X_{ph}$ which is plotted in black solid lines in figure \ref{fig:reyStress}. This helps to verify
whether the differences between subspace statistics and the full-field, are a result of the constraints mentioned in
chapter \ref{sec:limitations}, or a property of the flow represented by the corresponding subspace.  The invariants of
Reynolds stress tensor are also calculated for each subspace to provide a measure of how the entire tensor compares with
that of the snapshots. The first invariant being equal to the turbulent kinetic energy is already reported in the
previous chapter for each subspace. The remaining two are presented in this chapter. 

To ensure the accuracy and reliability of the simulated data, Reynolds stress components of the full-filed are plotted
in grey solid lines and are compared against the benchmark DNS data by \citet{el-khoury_schlatter_2013} plotted in red dashed lines in
figure \ref{fig:reyStress}.  Stress tensor components of the full-field agree very well with those of the benchmark with
the peaks being located at $y^+ = [15, 56, 36, 32]$ for $\langle u^2 \rangle$, $\langle v^2 \rangle$, $\langle w^2
\rangle $ and $\langle uv \rangle $ respectively. 

Subspace \rom{1} (s\rom{1}), plotted in solid blue lines, shows substantial contributions to the stress
components of the snapshots with average contribution of $98\%$. The wall-normal locations of the peaks coincide with
those of the snapshots at $y^+ = [14, 50, 28, 28]$ for $\langle u^2 \rangle$, $\langle v^2 \rangle$, $\langle w^2
\rangle $ and $\langle uv \rangle $. The second and third stress tensor invariants of this subspace amount to $97\%$ 
and $96\%$ of those of the snapshots.

Subspace \rom{2} which appears with axial length-scales smaller than s\rom{1} and larger than s\rom{3} (figure
\ref{ux3D}) is plotted in green solid lines. It contributes most to the radial stress component ($22\%$) and least to
the axial-radial one ($6\%$) reaching the peak values at wall-normal locations $y^+=28$ and $y^+=14$ respectively. The
peaks of axial and azimuthal components occur at $y^+=14$ and $y^+=16$ with $14\%$ and $21\%$ contributions to the
corresponding components of snapshots. Except for the axial component, all the peaks in this subspace have moved clearly
closer to the wall compared to the snapshots. The second and third invariants of the stress tensor of this subspace 
are equal to $2\%$ and $0.4\%$ of those of the snapshots respectively.

Subspace \rom{3} accommodating very small scale structures and represented
by $82\%$ of the modes has $3.3\%$ average contribution to the diagonal Reynolds stress components and $0.1\%$ to
$\langle uv \rangle$ reaching their maxima at $y^+ = [12, 92, 74, 31]$ respectively. This subspace contributes less than
$0.02\%$ to the second and third invariants of stress tensor.

%% file: spectra.tex
\subsection {Energy spectra}
\label{sec:spectra}

\begin{figure}
\centering
     {{ \includegraphics[width=0.48\linewidth]{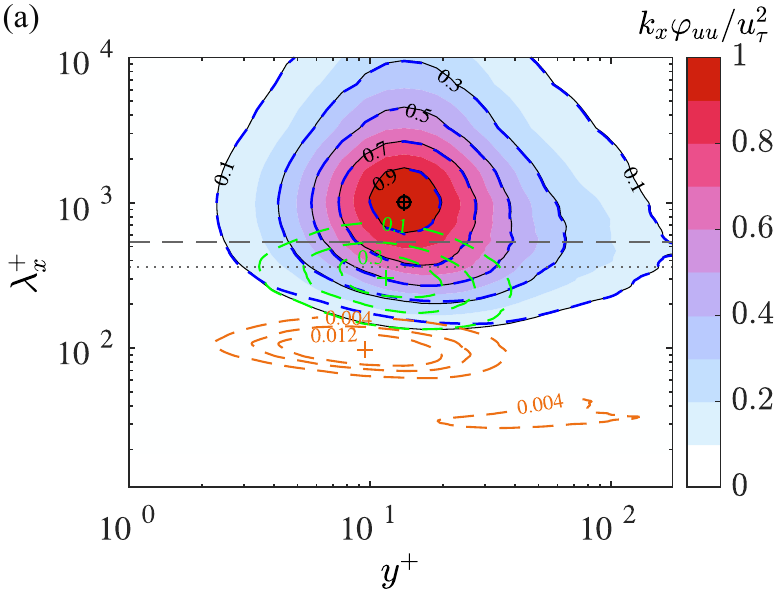} }}
     {{ \includegraphics[width=0.48\linewidth]{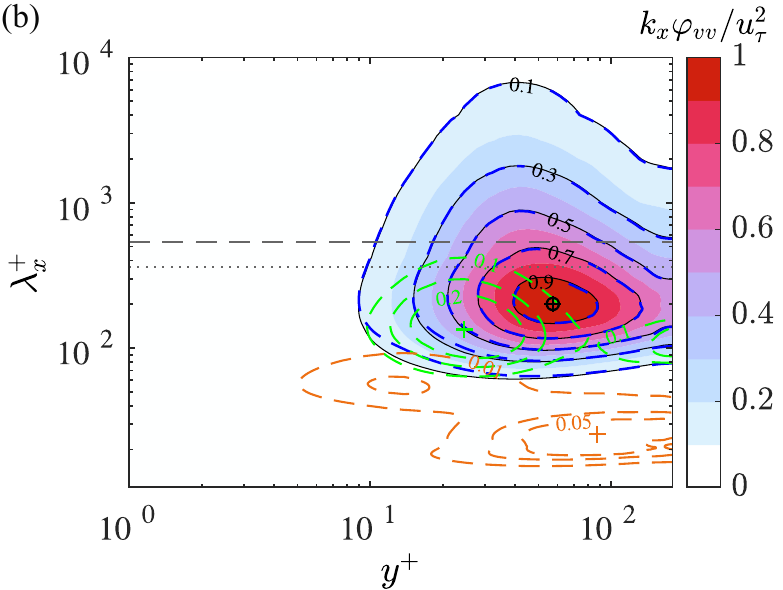} }}\\
     {{ \includegraphics[width=0.48\linewidth]{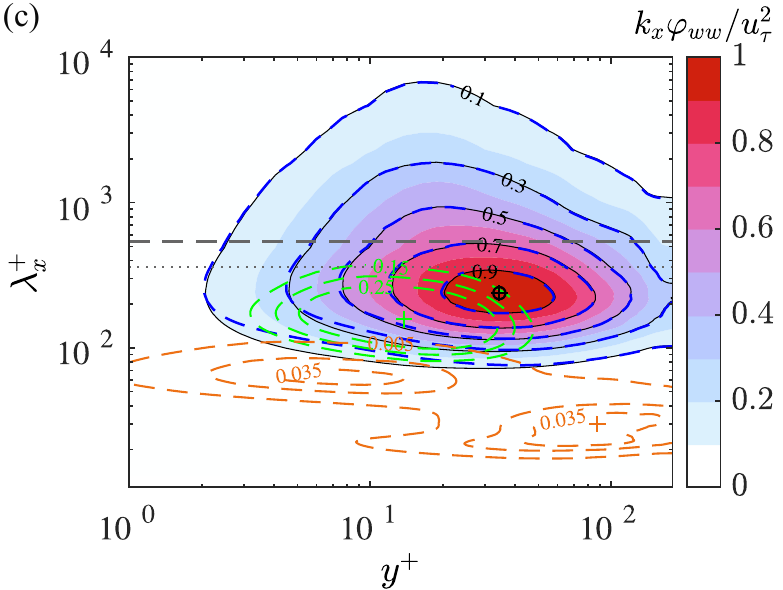} }}
     {{ \includegraphics[width=0.48\linewidth]{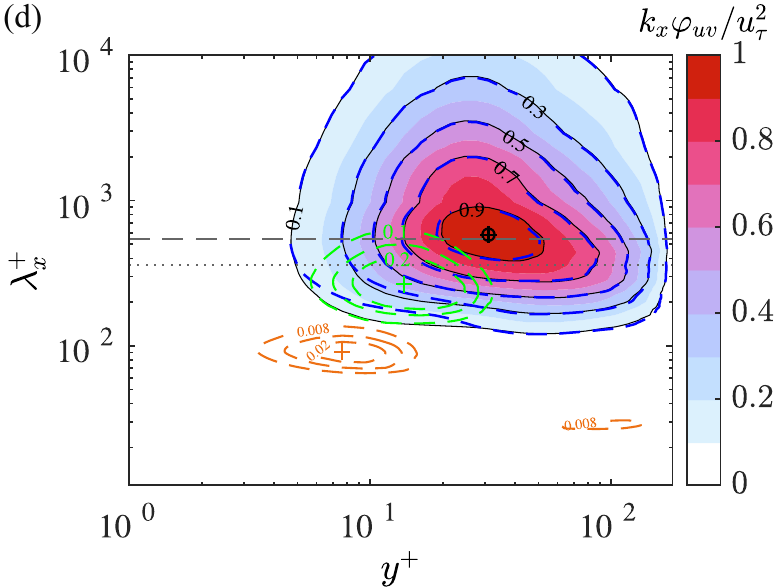} }}
    \caption{Premultiplied energy spectra of velocity auto correlations $\varphi_{uu}$ (a) ,
    $\varphi_{vv}$ (b), $\varphi_{ww}$ (c) and cross correlation $\varphi_{uv}$ (d).}
\label{KxPhi}
\end{figure}

The energy content of each length-scale is analysed for each subspace using premultiplied streamwise energy spectra of velocity auto
correlations ($ \varphi_{uu}, \varphi_{vv}, \varphi_{ww}$) and cross correlation ($ \varphi_{uv}$) plotted in figure
\ref{KxPhi} for the snapshots in coloured contours and black contour lines. Blue, green and orange dashed contour levels
represent subspaces \rom{1}, \rom{2} and \rom{3} respectively, each normalised by the maximum of the snapshots spectra.
Blue dashed lines and black solid contour lines correspond to the same levels annotated in black. Black circle and 
plus markers indicate spectral peaks of the snapshots and subspace \rom{1}. Coloured plus markers point to the 
peak locations of the corresponding subspace and coloured labels indicate the respective contour levels. The horizontal 
dotted and dashed lines are plotted as a reference for the commonly accepted axial length-scales of LSMs at $\lambda =
2R$ and $3R$ respectively. 

In the spectra of axial velocity in figure \ref{KxPhi}a, all large-scale structures are captured in
subspace \rom{1} and maximum energy is found for wave-length $\lambda^+=1006$ at $y^+=13.8$. Smaller
structures with up to length-scale of $\lambda^+=300$ are also present in this subspace. Energy of wave-lengths
$\lambda^+ \leq 300$ drop compared to the snapshots at $ 3 \leq y^+ \leq 30$, where the solid
black levels diverge from the dashed blue ones. Subspaces \rom{2} and \rom{3} appear with spectral peaks having
smaller axial length-scales of $\lambda^+=304$ and $97$ at $y^+ = 11.6$ and $9.5$ with normalised peak energy of $0.37$ and $0.02$ 
respectively.

The radial velocity component has the shortest axial wave-length compared to the other two components as seen in figure
\ref{KxPhi}b  with the main peak occurring at $y^+=57$ for $\lambda^+=201$ for the snapshots and s\rom{1}. The peaks of
subspaces \rom{2} and \rom{3} emerge with smaller wave-lengths of $\lambda^+=134$ and $25$ at $y^+ = 24$ and $87$ with
normalised peak energy of $0.3$ and $0.084$ respectively.

What can be observed in all subplots of figure \ref{KxPhi}, is that the spectral peaks found for subspace \rom{1}
coincide with those in the snapshots, in terms of their wall-normal locations and axial wave-lengths, with their energy
content peaking on average at $99\%$ of the snapshots spectral peaks.  Subspace \rom{1} has captured large-scale
energetic structures, and where its energy diverges from the snapshots, the next subspaces emerge with a peak. Spectral
peaks in subspace \rom{2} show a strong shift to the vicinity of the wall, although the shift is smaller for
$k_x \varphi_{uu}$. Subspace \rom{3} appears in all spectral maps with two low energy peaks, one below the main peak
closer to the wall and one above it, with length-scales $\lambda^+ \leq 100$.
The more energetic peak belongs to the one with larger wave-length for 
$\varphi_{uu}$ and $\varphi_{uv}$, whereas for $\varphi_{vv}$ and $\varphi_{ww}$ it represents the smaller wave-length.

%

%% file: isotropy.tex
\subsection{Anisotropy invariant map of the subspaces}
\label{sec:isotropy}

We study the structure of turbulent flow in each subspace by investigating the
invariants of anisotropic Reynolds stress tensor

\begin{equation}
    a_{ij} = \frac{\tau_{ij}}{\tau_{kk}} - \frac{\delta_{ij}}{3}.
\label{eq:anisoStress}
\end{equation}

\begin{figure}
\centering
    {{ \includegraphics{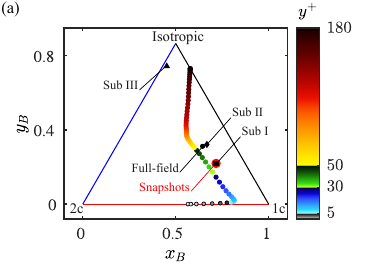} }}
    {{ \includegraphics{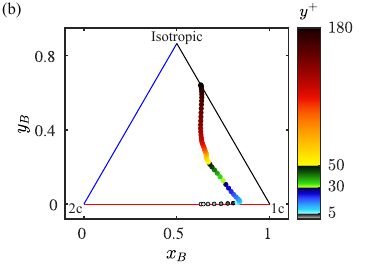} }} \\
    {{ \includegraphics{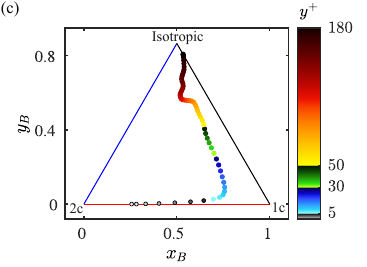} }}
    {{ \includegraphics{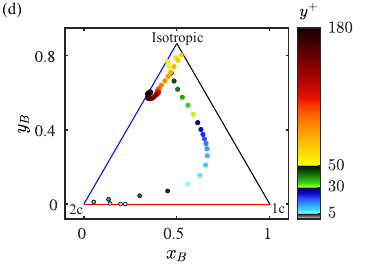} }}
\caption{Isotropy invariant map of the full-field (a) subspace I (b),
subspace II (c) and subspace III (d).}
\label{fig:AIM}
\end{figure}

A triangular domain is introduced by \citet{banerjee_2007} as a barycentric anisotropy map inside which  
all realisable Reynolds stress invariants are located.
The vertices of the triangle represent three limiting states of one-component (1c), two-component (2c) and three-component 
isotropic turbulence, located respectively
at $\textsc{x}_{1c} = (1,0)$, $\textsc{x}_{2c} = (0,0)$ and $ \textsc{x}_{3c} = (1/2, \sqrt{3}/2)$ as shown
in figure \ref{fig:AIM}. Moving away from the isotropic vertex on the blue edge corresponds to axi-symmetric contraction which 
ends up at the disc-like anisotropy at $2c$ vertex. Alternatively, moving on the black edge towards the $1c$ vertex corresponds to 
axi-symmetric expansion leading to needle-like anisotropy. The red edge connecting $1c$ and $2c$ vertices depicts 
the two-component limit. This map is defined using a linear combination of positive scalar metrics. These metrics
are functions of eigen values of $a_{ij}$ being sorted as $\lambda_1 \ge \lambda_2 \ge \lambda_3$ and are used to defined 
the coordinate system ($x_B,y_B$) given by

\begin{subeqnarray}
    x_B & = & C_{1c} x_{1c} + C_{2c} x_{2c} + C_{3c} x_{3c} = C_{1c} + C_{3c} \frac{1}{2} \, ,\\[3pt]
    y_B & = & C_{1c} y_{1c} + C_{2c} y_{2c} + C_{3c} y_{3c} = C_{3c} \frac{\sqrt{3}}{2}  \, ,
\end{subeqnarray}

\noindent where 

\begin{equation}
      C_{1c}  = \lambda_1 - \lambda_2 \, , \,\,\, C_{2c}  = 2 (\lambda_2 - \lambda_3) \,, \,\,\, C_{3c} = 3 \lambda_3 +1 \, .
\label{eq:AIMweights}
\end{equation}


To have a measure of the total anisotropy, $a_{ij}$ is calculated using temporal averaging and weighted spatial
averaging over all radial points and the results are plotted in annotated black markers in figure \ref{fig:AIM}a.  Subspaces
\rom{1}, \rom{2} and \rom{3} (shown with rectangle, diamond and triangle markers respectively) appear to be aligned on 
a line moving towards the isotropic state, with subspace \rom{3}
clearly being the most isotropic one. Snapshots averaged isotropy is plotted with a red circular marker being away from the
full-field due to the fewer number of timesteps available in the snapshots. Subspace \rom{1} is almost exactly on top
of the red marker showing a similar anisotropy to the snapshots. 

The invariant map is plotted for the full-field and for each wall-normal location in figure \ref{fig:AIM}.  To clearly
inspect the isotropy state at each wall layer, a different colourmap has been chosen to distinguish four wall layers. Grey
colourmap is set for the viscous sublayer, blue for the buffer layer between the viscous and logarithmic layer ($5 \le
y^+ \le 30$), green for the logarithmic layer and a heat colourmap is chosen for the overlap and outer layer ($50 \le y^+
\le 181$). The map starts for the full-field at the wall at the two-component limit in figure \ref{fig:AIM}a and moves towards the
one-component vertex. At $y^+ = 10$ in the buffer layer a sharp bend is observed after which the trajectory moves
towards the centre of the map where a second bend is reached at $y^+ \approx 83$ followed by a straight path towards
the isotropic vertex at the centre of the pipe. 

The map for subspace \rom{1} is plotted in figure \ref{fig:AIM}b for each wall-normal location. The first bend takes
place at the same location as the full-field whereas the second bend appears earlier at $y^+=65$ followed by an S shaped
movement towards the isotropic state. Similarly, subspace \rom{2} starts on the wall on the two-component limit but
closer to disc-like isotropy moving more rapidly towards the $1c$ vertex, reaching a softer bend at $y^+ = 10$. After
that, the trajectory follows a straight line approaching the axi-symmetric expansion limit where the second bend takes
place moving away from the black edge at $y^+ \approx 83$. A third bend is reached at $y^+ \approx 120$ after which the
flow approaches the isotropic state close to the pipe axis. 

Subspace \rom{3} shows a very different behaviour starting on the two-component limit and moving towards the disc-like
anisotropy where it almost reaches the $2c$ vertex in the viscous sublayer at $y^+=1.3$. A very soft bend takes place
at $ 7.5 \le y^+ \le 14 $ followed by a path towards the isotropic state. In the overlap layer at $ 57 \le y^+ \le 65$ the
trajectory gets closest to the isotropic state after which it departs towards the axi-symmetric contraction limit at the
pipe axis.

%% file: summary.tex
\subsection{Summary}

A characteristic DMD is carried out on DNS data of turbulent pipe flow at $\Rey_b = 5300$ decomposing three velocity
components along the characteristics of the flow corresponding to the group velocity of $u_g=1.1\, U_b$.  Three
subspaces are extracted and their contributions to stream wise energy spectra and components of Reynolds stress tensor
are investigated along with the anisotropy invariant maps to compare the structure of turbulence in each subspace.

Subspace \rom{1} being the most energetic one is comprised of 11 modes ($3\%$ of the total modes) and is detected based
on three main criteria: the mode amplitudes, cumulative energy of the constituent modes and the relative error with
respect to the snapshots matrix. This subspace undergoes minimal oscillations in space and time having a very small
average frequency of $\hat{f_I} = 0.086$ and it decays slowly with the mean decay rate of $\hat{d_I} = 0.022$. The modes in
this subspace form small angles with one another and larger ones with the rest of the modes, indicating large energy
interactions inside the subspace and small interactions with the rest of the modes while maintaining the total kinetic
energy of the subspace. The axial wave lengths and wall normal locations of the spectral peaks in this subspace coincide accurately 
with those of the snapshots. Subspace \rom{1} contributes $97\%$ to the turbulent kinetic energy and $99\%$ to the
$\langle uv \rangle$ component of Reynolds stress tensor. 

Subspace \rom{2} is detected with 67 modes ($15\%$ of the total modes) based on cumulative energy interactions of its
members to subspace \rom{1}. This subspace oscillates almost 25 times faster than subspace \rom{1} with average
frequency of $\hat{f}_{II}=2.1$ and decays almost twice faster with average decay rate of $\hat{d}_{II}=0.043$. The
spectral peaks of s\rom{2} in all axial energy spectra appear closer to the wall in the buffer layer with the peak
value amounting to $30\% - 40\%$ of the snapshots peak. Only small scale flow features are present in this subspace with
spectral peaks corresponding to maximum wave length of $\lambda^+=304$ for the stream wise component and minimum of
$\lambda^+=134$ for the radial one. The peaks of Reynolds stress profiles emerge closer to the wall for all the
components having total contribution of $6\%$ to $\langle uv \rangle$ component of Reynolds stress tensor. This subspace
contributes $15\%$ to kinetic energy while its interactions with subspaces \rom{1} and \rom{3} causes $12\%$ and $2\%$
of energy loss respectively. There are length scales which are present in s\rom{1} and s\rom{2}
implying that all structures with the same length scales do not necessarily have the same contributions to
Reynolds stress tensor or kinetic energy.

The remaining 346 modes ($82\%$) constitute subspace \rom{3} having only minimal energy contributions to the first two
subspaces. It oscillates on average more than two times faster than subspace \rom{2} with frequency of $\hat{f}_{III} =
5.61$ and decays with average decay rate of $\hat{d}_{III}= 0.016$. Only very small scale flow features with low energy
levels are observed here which are oscillating fast but they are persistent in space and time. This subspace has $2\%$
contribution to the kinetic energy and close to no overall interaction with subspace \rom{1}. It contributes to $0.1\%$
of $\langle uv \rangle$ and shows the most isotropic behaviour among all the subspaces specially in the overlap layer
and a distinct disc-like anisotropy in the viscous sublayer. 

\subsection{Conclusions}
The wave-like definition of coherent structures in a characteristic frame of reference in transport-dominated turbulent
flows proves to be very efficient for the two following reasons: The main features of pipe flow at $Re_b=5300$ is
captured accurately with only $3\%$ of the modes which form an almost orthogonal subspace to the rest of the modes.
This subspace reproduces $97\%$ of the turbulent kinetic energy of the full flow, and more than $96\%$ of the invariants
of the Reynolds stress tensor. Its spectral signature matches the snapshots in terms of wall-normal locations and
wavelengths of the premultiplied energy spectra.

The second reason is that the remaining modes can be further divided into two subspaces based on their cumulative
energy contributions to subspace \rom{1}. The Third subspace accommodates very small scales with
short turn-over times and persists as a turbulent background motion. The second subspace lives in between the
mentioned subspaces \rom{1} and \rom{3} having faster decay rates than the other two subspaces with their spectral peak
length scales being substantially smaller than the first and substantially larger than the third subspace.

We speculate that at higher Reynolds numbers more modes would be needed to build subspace \rom{1} and the scale
separation between the subspaces would increase. We base our speculation on the fact that the flow becomes more complex 
and wider range of group velocities would be present in the flow.